\documentclass{jicspack}
\usepackage{graphicx}
\usepackage{caption}
\usepackage{subcaption}

\setvolume{10}  
\setyear{2013}                        
\setpagerange{1}{8}                   
\setheadauthor{X. Liu et al.}          
\setissn{1548--7741}%
\setpubdate{January 2013}%
\setno{1}     

\setdoi{10.12733/jicsxxxxxxxx}

\afterpage{\beginheader}                   

\begin{document}
\begin{premaker}

\title{Finite Element Method Based Modeling of Cardiac Deformation Estimation under Abnormal Ventricular Muscle Conditions}

\author[author1]{Ahmadreza Baghaie\corauthref{cor1}},
\ead{abaghaie@uwm.edu}
\corauth[cor1]{Corresponding author.}
\author[author2]{Hamid Abrishami Moghaddam}

\address[author1]{Department of Electrical Engineering, University of Wisconsin-Milwaukee, WI, USA 53211}
\address[author2]{Department of Biomedical Engineering, K.N.Toosi University of Technology, Tehran, Iran}


\begin{abstract}
Deformation modeling of cardiac muscle is an important issue in the field of cardiac analysis. Many approaches have been developed to better estimate the cardiac muscle deformation, and to obtain a practical model to be used in diagnostic procedures. But there are some conditions, like in case of myocardial infarction, in which the regular modeling approaches are not useful. In this article, using a point-wise approach, we try to estimate the deformation under some abnormal conditions of cardiac muscle. First, the endocardial and epicardial contour points are ordered with respect to the center of gravity of the endocardial contour and displacement vectors of boundary points are extracted. Then to solve the governing equations of the deformation, which is an elliptic equation, we apply boundary conditions in accordance with the computed displacement vectors and then the Finite Element method (FEM) will be used to solve the governing equations. Using the obtained displacement field of the cardiac muscle, strain map is extracted to show the mechanical behavior of the cardiac muscle. Several tests are conducted using phantom and real cardiac data in order to show the validity of the proposed method. 
\end{abstract}
\begin{keyword}
Cardiac Modeling, Myocardial Infarction, Finite Element Method (FEM)
\end{keyword}
\end{premaker}


\section{Introduction}\label{Intro}

Development of a consistent deformation estimation model for both accurate estimation
of cardiac deformations and taking into account of the organ's regional abnormalities is an interesting area of research in biomedical engineering. On the other hand, due to the diagnostic importance of such analysis and the inherent restrictions of medical instruments in data acquisition, new approaches are proposed to facilitate the interpretation of the acquired data. Low contrast and noisy cardiac images provided by medical imaging systems are examples of the restrictions in biomedical devices. To overcome the above limitations,
medical data analysis methods use \textit{a priori} knowledge about organ's behavior. These model-based approaches can achieve more accurate and robust results. In the model based analysis, using natural properties of the organ, the performance of algorithms are improved.

There are some other limitations in such analysis which in case of cardiac modeling
will become important. On one hand the importance of such analysis because of its
association with human health and on the other hand the difficulties of modeling, caused by
a wide range of cardiac diseases and also the device limitations, play a critical role in these types of researches. Also there are several syndromes caused by different coronary heart diseases, with different interpretation and effect on cardiac features. Among them, Myocardial Infarction (MI) can cause serious problems in deformation
based cardiac modeling approaches. On one hand, existence of infarction in cardiac
muscle may cause some abnormalities in cardiac deformation's expressing equations. On
the other hand, these areas should be recognized and located to facilitate the procedure of
cardiac deformation modeling. In this manner, in recent years, some approaches have been
introduced to extend the previous models to overcome such problems. 

In 1993, Huang et al, using an adaptive mesh size, analyzed cardiac deformation \cite{1}.
McInerney and Trezopolos used a dynamic balloon model to extract the endocardium
surface \cite{2}. In the model, the balloon deforms until it is laid properly on the endocardium
surface. Faber also used a deformable template for 3D demonstration of heart's left
ventricle \cite{3}. Also in the same year, snake model is used for 2D segmentation of cardiac
muscle \cite{4}. In \cite{5} a deformable template is used for deformation analysis of cardiac muscle in
2D images and in \cite{6}, Amini et al used snake models for tracking cardiac wall. Active mesh
model is used in the work of Rukert et. al. for examination of cardiac deformation \cite{7}. In 1998,
using Tagged MRI images and snake model, 2D deformation of cardiac muscle is analyzed
\cite{8}. Also in the same year, snake models are used for analysis of the cardiac
muscle \cite{9} and for dynamic demonstration of cardiac muscle in one cardiac cycle. Shi et al, proposed Delaunay active mesh for point to point tracking of
heart's left ventricle \cite{10}.

In neither of the above, cardiac abnormality is
considered. In this case, the model based approaches were unable to detect and localize
infarct zone in cardiac muscles. In 2001, using Elastography, local ranges of cardiac strain was computed and was used to
distinguish between normal and inefficient cardiac muscle \cite{11}. In \cite{12}, the Post
Systolic Stretching (PSS) criterion of the cardiac muscle was used for this purpose. In \cite{13}, a new \textit{in-vivo}
statistical model for estimation of the material properties and strain/stress distribution in cardiac muscle for both right and left ventricles is proposed. The results of applying the
method on normal and right ventricular hypertrophy (RVH) cases, demonstrate a good performance in modeling of cardiac muscle. In 2004, using CT imaging system, the attenuation rate of infarct region and normal regions were compared and based on this, the size of the infarct zone was computed \cite{14}. In the same year, by determining end-systolic volume, end-diastolic volume, stroke volume and etc, operational differences between normal and
abnormal cardiac muscle were derived \cite{15}.

In \cite{16}, a semi-automatic computer interface is used for infarct localization and based on a
threshold signal density, the infarct zone divided to core-infarct zone and neighbor-infarct
zone. In \cite{17}, an analysis based on the cardiac strain model derived from tissue Doppler imaging
of normal and abnormal cardiac muscles is done. In the same year, a 4-dimensional
model is used for mice cardiac muscle contraction operation in ultrasonic images and the
model was applied to both normal and infarct zone cardiac muscles \cite{18}. In \cite{19} using a
normalized parametric domain, normal and abnormal cardiac muscle were modeled. In our
work [20], an appropriate model is proposed for analysis of infarct zone cardiac muscles.
Here, the proposed model in \cite{20} is extended to determine the strain ranges of the cardiac
muscle in normal and abnormal conditions. Moreover, localization of the infarct zone in
cardiac muscle is done using a well-known criterion, namely \textit{Effective Stress}, in mechanical analysis.

The article is organized as follows. In Section 2  dynamic modeling of the cardiac muscle will be investigated and the governing equations, the boundary conditions and the Finite Element model will be introduced. Section 3 contains the results of applying the proposed model on phantom data. Modeling and analysis of the real cardiac data and the analysis of the left ventricle's volumetric changes in case of infarction is presented in Section 4. Section 5 concludes the paper.

\section{Dynamic Modeling of the Cardiac Muscle}
\subsection{Elastic Model}
Assume $\{\sigma\}$ is the stress array and $\{\varepsilon\}$ is the strain array. Also assume that $[E]$ is the
structural matrix consisting of elastic coefficients in the model. For a linear elastic condition,
stress-strain equation can be represented in matrix form as:
\begin{equation}
\{\sigma\}=[E]\{\varepsilon\}+\{\sigma_0\}
\end{equation}
where $\{\sigma_0\}$ is the array of initial stress which will be considered here as 0. For 2D case, matrix $[E]$ is a $3\times 3$ matrix. This matrix is symmetric $E_{ij}=E_{ji}$. For an isotropic material and in plane strain condition, the matrix can be written in the following form \cite{21}:
\begin{equation}
[E]=\frac{E}{1-\nu^2} \begin{pmatrix}
  1 & \nu & 0 \\
  \nu & 1 & 0 \\
  0 & 0 & \frac{1-\nu}{2}
 \end{pmatrix}
\end{equation}
where $\nu$ is the Poisson's ratio and $E$ is the Young's modulus. Considering $u$ and $v$ as displacements in $x$ and $y$ directions respectively, the strain-displacement
equation can be expressed as \cite{22}: 

\begin{equation}
\begin{pmatrix} \epsilon_x \\ \epsilon_y \\ \gamma_{xy} \end{pmatrix}
=\begin{pmatrix}     

\frac{\partial}{\partial x} & 0 \\
0 & \frac{\partial}{\partial y} \\
\frac{\partial}{\partial y} & \frac{\partial}{\partial x}
\end{pmatrix}
\begin{bmatrix}
u \\ v
\end{bmatrix} 
= [\boldmath{\partial}]\{\textbf{u}\}
\end{equation}
where $\{\textbf{u}\}$ is the displacement vector. 

In general, when an object deforms, if there is no breakdown, for satisfying the consistency
condition, it is necessary that the displacements are continuous and single-valued functions
of the object's position. On the other hand, every deforming object needs to satisfy the
equilibrium equation. The equilibrium equation is \cite{22}: 
\begin{equation}
[\boldmath{\partial}]^T \{\boldmath{\sigma}\}+\{\textbf{F}\}=\{\textbf{0}\}
\end{equation}
where $\{\textbf{F}\}$ is a vector consisting of $F_x$ and $F_y$ elements, which are body forces in $x$ and $y$
directions respectively and $\{\textbf{0}\}$ is a vector with elements equal to 0. Another constraint for deformation is the boundary conditions which are the displacement or strain on the face or surface of the deforming object.

\subsection{Displacement Computation on the Boundary}
For modeling of the behavior of the deforming object, it is necessary that the displacement
filed is determined through the object. For this, assume that $X(0)$ is the initial coordinates of a point in the object in $t=0$ and $X(t_0)$ is the coordinates of the same point at $t=t_0$. With this assumption, the displacement vector between these two time steps can be defined as: 
\begin{equation}
U=X(t_0)-X(0)
\end{equation}

The problem is that defining a one-to-one correspondence between points of an object is a
difficult task. In the problem at hand, dynamic modeling of cardiac muscle, the main goal is to determine the
inter-wall displacement field between two time steps. Then using this field, the strain map
will be computed. So it is necessary to define a one-to-one correspondence between body
points in two different times. 

Suppose that $I(0)$ and $O(0)$ are matrices containing inner and outer contour points of the
cardiac muscle at $t=0$ respectively and $I(t_0)$ and $O(t_0)$ are matrices containing inner
and outer contour points at $t=t_0$. Also assume that $[x_c, y_c]^T$ represents the coordinates
of the center of gravity of inner contour at $t=0$. With these assumptions the coordinates
of all contour points, inner and outer, can be represented with respect to the center point. In
other words, for simplicity, the position of the center of gravity can be assumed fixed and the
contour points' positions are represented with respect to the center point as the center of
coordinate system. It should be mentioned that this assumption is only valid for deforming
objects with small strain which is the case of analysis in cardiac modeling. Also the cardiac
muscle deformation is a radial deformation and this assumption is not valid for non-radial
deformations. This assumption can lead to increased error in the deformation analysis of non-convex objects. 

Based on the above mentioned notes, the displacement vectors can be computed in the
following order:
\begin{enumerate}
\item The center of gravity of the inner contour points is computed and assumed fixed
from time $t=0$ to time $t=t_0$.
\item In accordance with the center, contour points can be ordered based on the angle of the crossing line between the point and the center and horizontal axis. Fig. 1
shows the procedure.
\item After ordering all the contour points, the displacement vectors for corresponding
points between two time steps (the points with the same angle) can be computed. 
\end{enumerate}

\begin{figure}
\centering
\includegraphics[scale=.6]{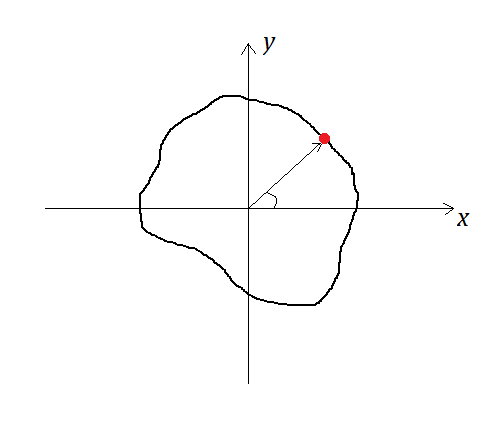}
\caption{Position of a boundary point with respect to the center of gravity of the contour}
\end{figure}

\subsection{Governing Equations of the Deformation}
Assume that $\Omega$ is the deforming body's domain and $\partial \Omega$ is the domain's boundary. During the deformation of the body, it is necessary that an equation consistent with the 3 aforementioned conditions is satisfied. The governing equation for deformation of a deformable body is an elliptic equation like following: 
\begin{equation}
-\bigtriangledown(c(u).\bigtriangledown u)+a(u)u=f(u) \quad on \quad \Omega
\end{equation}
where $f$, $a$ and $c$ are functions of unknown $u$ which is to be solved. These functions have complex values in the domain $\Omega$. By expanding this equation we have: 
\begin{equation}
\begin{split}
-\bigtriangledown(c_{11}. \bigtriangledown u_1)- \bigtriangledown(c_{12}. \bigtriangledown u_2)+a_{11}u_1+a_{12}u_2=f_1 \\
-\bigtriangledown(c_{21}. \bigtriangledown u_1)- \bigtriangledown(c_{22}. \bigtriangledown u_2)+a_{21}u_1+a_{22}u_2=f_2
\end{split}
\end{equation}

Dirichlet and Neumann boundary conditions for scalar $u$ can be defined as: 
\begin{equation}
\begin{split}
hu=r \quad on \quad \partial \Omega \\
\overrightarrow{n}.(c\bigtriangledown u)+qu=g \quad on \quad \partial \Omega
\end{split}
\end{equation}
where $\overrightarrow{n}$  is the outward normal vector, orthogonal to the boundary. $g$, $q$, $h$ and $r$ are functions with complex values on the boundary $\partial \Omega$. For a 2D system, the Dirichlet boundary condition is: 
\begin{equation}
\begin{split}
h_{11}u_1+h_{12}u_2=r_1\\
h_{21}u_1+h_{22}u_2=r_2\\
\end{split}
\end{equation}

The Neumann boundary condition can be achieved in the same manner. Now it is the time to determine the boundary conditions for solving the governing equations. Fig. 2 displays two adjacent contour points, $i$ and $j$, with their corresponding points after the displacement. The displacement vectors for these two points are: 
\begin{equation}
\begin{split}
\{u\}_i=\begin{pmatrix}
u_i \\ v_i
\end{pmatrix} \\
\{u\}_j=\begin{pmatrix}
u_j \\ v_j
\end{pmatrix}
\end{split}
\end{equation}

\begin{figure}
\centering
\includegraphics[scale=.6]{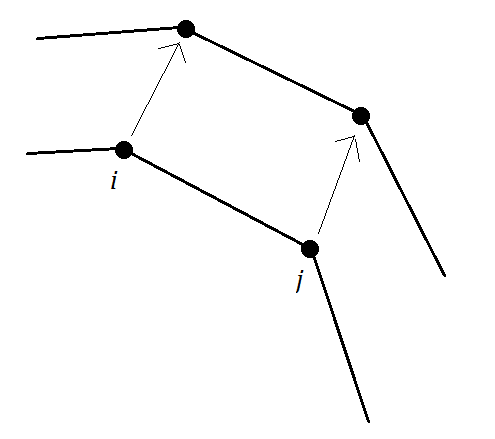}
\caption{Boundary nodes, $i$ and $j$, and their corresponding points after deformation}
\end{figure}

Here we only consider the displacements in $x$ and $y$ directions, so $h_{12}=h_{21}=0$. Therefore for boundary conditions of the boundary line consisting $i$ and $j$ contour points: 
\begin{equation}
\begin{split}
h_{11}u_{ii}=\frac{u_i+u_j}{2}\\
h_{22}u_{jj}=\frac{v_i+v_j}{2}
\end{split}
\end{equation}

Boundary conditions for other boundary lines consisting of any two adjacent contour points
will be computed in the same manner. After determining the boundary conditions, and
producing triangular meshes, the problem is ready to be solved by Finite Element Method (FEM).
In FEM first the domain is divided into simple geometrical shapes, like triangles, and then
based on the assumption that the solution in the domain of each division, or element, has a
simple form, the whole domain's solution is computed. Polynomial approximation functions are useful choices for approximation of the solution in each element. The evaluation of these functions is simple they produce good approximates on small elements. It should be mentioned that in FEM one of the basic assumptions is that the solution in the domain is continuous. Therefore, it is necessary to have identical results on the shared sides of elements.

Suppose that $u_h$ is a continuous piece-wise linear approximation of $u$ in the elliptic equation. The goal is finding the best approximation for $u$ from polynomial functions class. So the function $u_h$ is compared with all of polynomial functions, $V$. We have: 
\begin{equation}
\int_\Omega (-\bigtriangledown(c.\bigtriangledown u_h)+a u_h-f)V d\Omega=0
\end{equation}
Therefore, based on the Green's theorem:
\begin{equation}
\int_\Omega (c\bigtriangledown u_h)\bigtriangledown V+a u_h V d\Omega - \int_{\partial \Omega} \overrightarrow{n}.(c \bigtriangledown u_h) V ds=\int_\Omega f V d \Omega \quad \forall V
\end{equation}

The boundary conditions are included in this equation. In general, every continuous piece-wise linear function can be written in the form: 
\begin{equation}
u_h(x,y)=\sum _{i=1} ^ N U_i \Phi_i (x,y)
\end{equation}
where $\Phi_i $'s are some continuous piece-wise linear basis functions and $U_i$'s
are scalar coefficients. $\Phi_i $'s are defined is such a way that they have value 1 in node $i$ and value $0$ on other nodes. For every function $V$, formulation of the finite element problems leads to an algebraic equation based on $U_i$'s. With the definition of $V=\Phi_j$ for $j = 1,2,..,,N$ where $N$ is the number of nodes defined in the domain, a system of equations in the form $KU = F$ can be achieved in which $K$ and $F$ consist of integral values related to $\Phi_i$ and $\Phi_j$ functions and problems coefficients and $U$ is the matrix of unknown scalar values in the defined nodes. For further explanation of the method, please refer to [22]. 

\subsection{Strain Computation in Triangular Elements}
Suppose that we have a triangular element with coordinates $[x_1, y_1]^T$, $[x_2, y_2]^T$ and $[x_3, y_3]^T$ as its nodes. Transforming point 1 to center of coordinates system in a manner that the line between points 1 and 2 lies on the positive side of $x$ axis, we have these transformed coordinates: $[0,0]^T$, $[x_2^{'}, 0]^T$ and $[x_3^{'}, y_3^{'}]^T$. With the assumption of basis functions as linear, two components of deformation can be written as:
\begin{equation}
u=\begin{pmatrix}
1 & x & y
\end{pmatrix} \begin{pmatrix}
a_1 \\ a_2 \\ a_3
\end{pmatrix}
, \quad
v=\begin{pmatrix}
1 & x & y
\end{pmatrix} \begin{pmatrix}
a_4 \\ a_5 \\ a_6
\end{pmatrix}
\end{equation} 
considering the strain-stress equation, $\varepsilon_x=a_2$, $\varepsilon_y=a_6$ and  $\gamma_{xy}=a_3+a_5$. Therefore defining $[u_1, v_1]^T$, $[u_2, v_2]^T$ and $[u_3, v_3]^T$  as displacement vectors of points 1, 2 and 3 respectively, three components of strain can be defined as: 
\begin{equation}
\begin{pmatrix}
\varepsilon_x \\ \varepsilon_y \\ \gamma_{xy}
\end{pmatrix} = \begin{pmatrix}
\frac{-1}{x_2^{'}} & 0 & \frac{1}{x_2^{'}} & 0 & 0 & 0 \\
0 & \frac{x_3^{'}-x_2^{'}}{x_2^{'}y_3^{'}} & 0 & \frac{-x_3^{'}}{x_2^{'}y_3^{'}} & 0 & \frac{1}{y_3^{'}} \\
\frac{x_3^{'}-x_2^{'}}{x_2^{'}y_3^{'}} & \frac{-1}{x_2^{'}} & \frac{-x_3^{'}}{x_2^{'}y_3^{'}} & \frac{1}{x_2^{'}} & \frac{1}{y_3^{'}} & 0
\end{pmatrix} \times \begin{pmatrix}
u_1 \\
v_1 \\
u_2 \\
v_2 \\
u_3 \\
v_3
\end{pmatrix}
\end{equation}
where $\varepsilon_x$, $\varepsilon_y$ and $\gamma_{xy}$  are triangular element's strain in $x$ and $y$ directions and shear strain respectively.

\section{Modeling of an Inhomogeneous Ring under Internal Radial Pressure}

\subsection{ANSYS Based Modeling }

At first it is necessary to verify the proposed model. For this, the results of the model should
be compared with the results of a valid experiment. To achieve a valid reference result for
comparison, ANSYS, well-known software in mechanical analysis, is used for basic modeling
of the problem. For modeling the left ventricle’s wall, many different models can be
introduced. Here we try to model the heart muscle with infarctions, which limits our model.
On the other hand determining the actual values of the parameters of heart muscle is a
different task. Therefore finding a proper model for verification of the proposed model can
be challenging, especially if we want to compute the stress map of the heart muscle. Here we only consider finding the strain map of heart muscle. 

In case of plane strain, the problem at hand can be simplified as follows. Assume that we
have an inhomogeneous ring cylinder with a known height. Since the foundation of our
research is based on the assumption of plain strain/stress, there is no strain/stress along the $z$ direction and therefore the cylinder's height doesn't have any effect on the solution. The assumption of an inhomogeneous ring cylinder is because of the changes
occurred in the mechanical properties of heart muscle after introducing infarction.
After defining the elastic properties of the ring cylinder, the structure is ready for the next step which is loading. In this step the load is applied in a uniform and radial manner inside the ring, to mimic the process of loading the heart muscle. The internal pressure applied to the ring is time varying. Fig.3 represents the results of ANSYS analysis of the ring for displacement in $x$, $y$ and the absolute values of displacements respectively. 

\begin{figure}
\centering
\begin{subfigure}[b]{.45\textwidth}
\includegraphics[scale=.35]{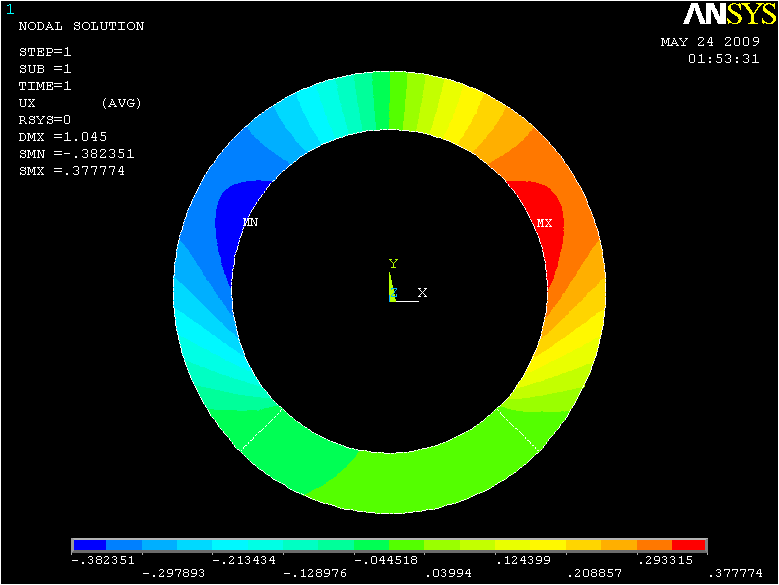}
\caption{}
\end{subfigure}
\begin{subfigure}[b]{0.45\textwidth}
\includegraphics[scale=.35]{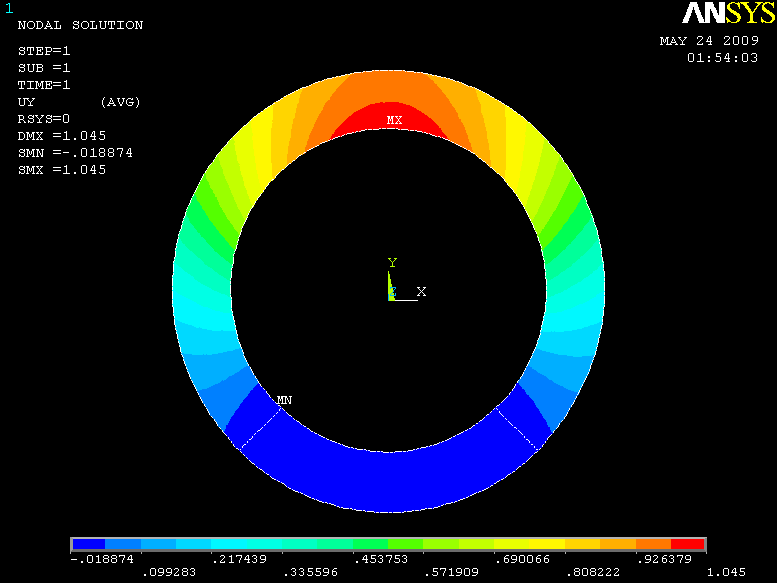}
\caption{•}
\end{subfigure}

\begin{subfigure}[b]{0.45\textwidth}
\includegraphics[scale=.35]{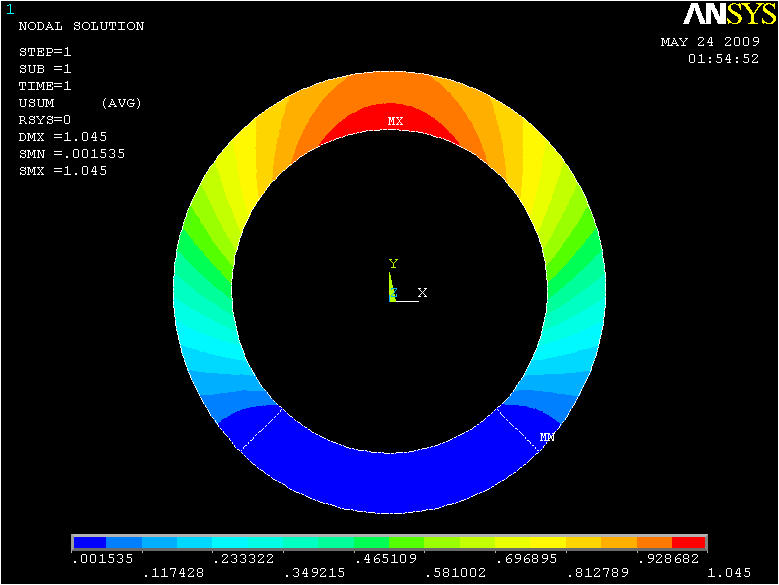}
\caption{•}
\end{subfigure}
\caption{ANSYS results for the displacement maps in $x$ (a) and $y$ (b) directions and the absolute displacement (c)}
\end{figure}

\subsection{Modeling With the Proposed Model}
After the analysis with ANSYS, the proposed model should be implemented and the results
should be compared and verified with the results of ANSYS analysis. At first the inner and our contours of the ring is determined. Then, after setting the boundary conditions, the governing equation of deformation is solved using FEM and the deformation maps are created. All the code is implemented in MATLAB. 

Fig. 4 represent the results of the proposed method for both horizontal, vertical and absolute displacement. To better be able to compare the results, the ring is divided into 16 sub-regions and the results of ANSYS and MATLAB implementations are compared. Fig. 5 shows the average displacement in each region for 10 iterations of pressure increment for both ANSYS and proposed method. Based on the provided results, it can be seen that the proposed model is consistent with the results of ANSYS, even in regions with low mobility (regions 11, 12, 13 and 14). 

\begin{figure}
\centering
\begin{subfigure}[b]{.4\textwidth}
\includegraphics[scale=.4]{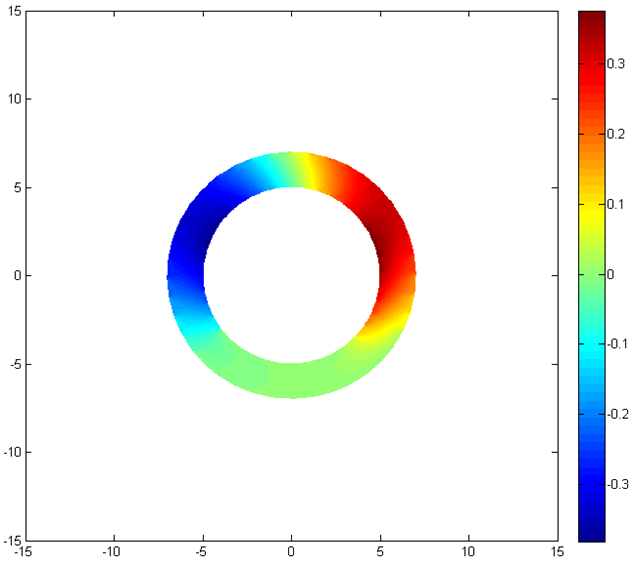}
\caption{}
\end{subfigure}
\begin{subfigure}[b]{0.4\textwidth}
\includegraphics[scale=.4]{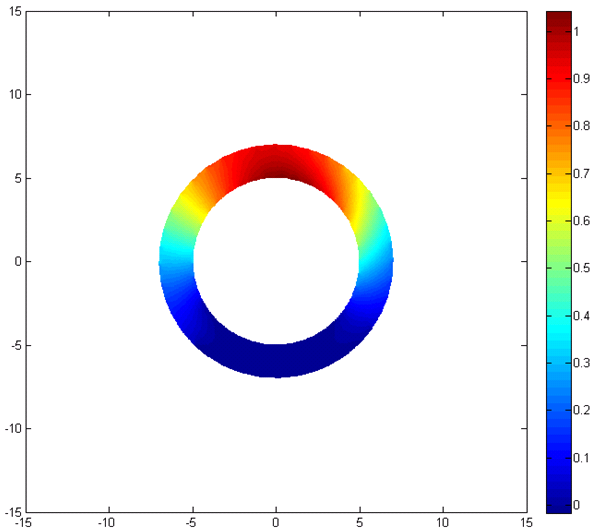}
\caption{•}
\end{subfigure}

\begin{subfigure}[b]{0.4\textwidth}
\includegraphics[scale=.4]{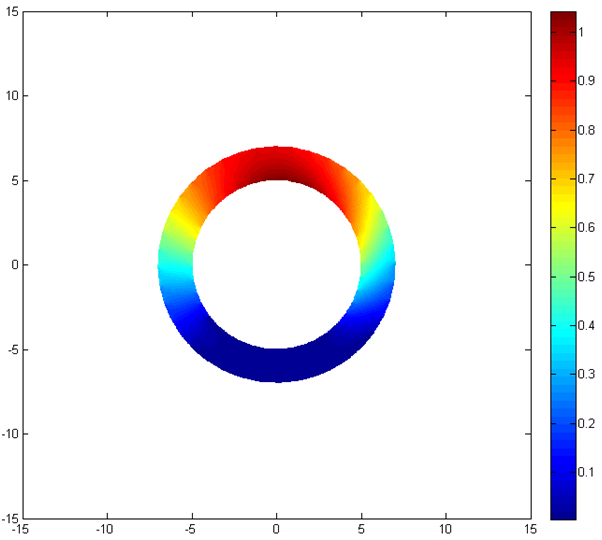}
\caption{•}
\end{subfigure}
\caption{Proposed model's results for the displacement maps in $x$ (a) and $y$ (b) directions and the absolute displacement (c)}
\end{figure}

\begin{figure}
\centering
\includegraphics[scale=.6]{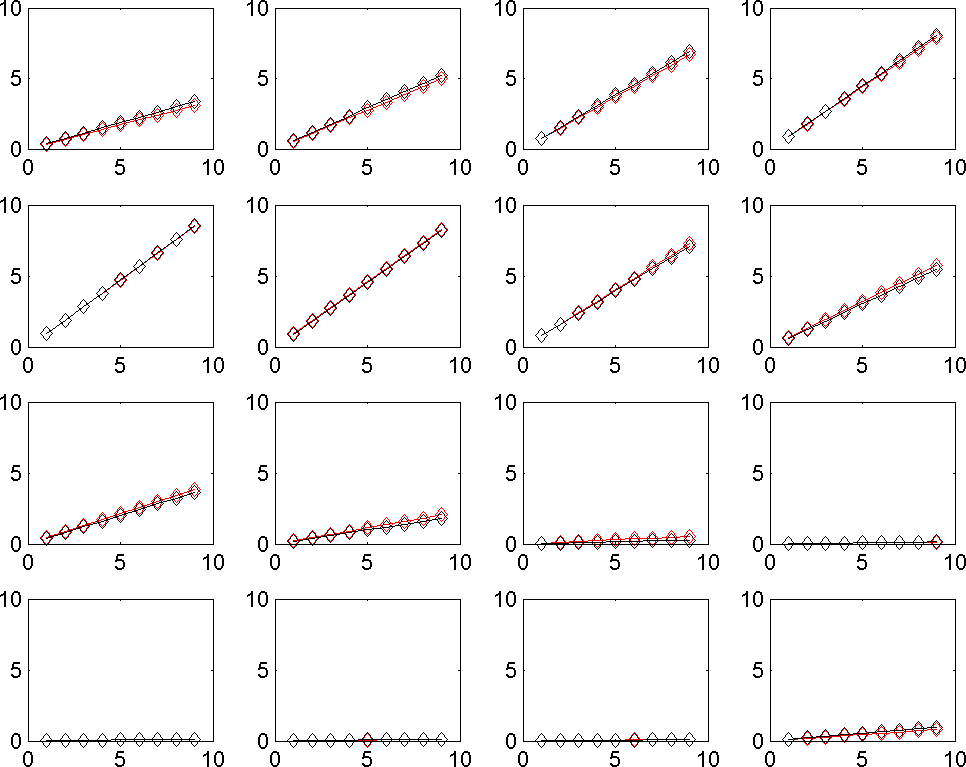}
\caption{Average displacement for ANSYS (red) and the proposed method (black)}
\end{figure}

\section{Modeling and Analysis of Heart Images Using the Proposed Algorithm }

\subsection{Database}

In this research the database provided by the authors of \cite{23} is used. This database
contains the MRI images of 33 patients which consists of 7980 two dimensional MRI images
from different slices of the heart muscle. Each cardiac cycle is divided into 20 steps. These
images are acquired from patients with different types of problems like Cardiomyopathy,
Aortic regurgitation, Ischemia and Enlarged ventricles. All of patients were under 18 and the
distance between slices is between 6 to 13 millimeters. Each image's size is $256\times256$ and the
inner and outer contour of the heart is indicated by hand by the author of \cite{23}. Based on the
additional information attached to the database it is known that patient 23 suffers from
Myocardial Infarction (MI). Also patients 18 and 25 have normal heart muscle which can be useful in understanding the effects of MI. Here the data from these three patients are used
to achieve the results. Fig. 6 displays the cardiac cycle of the patient 23 for the first slice. The inner and outer contours of the left ventricle are determined in the images. 

\begin{figure}
\centering
\includegraphics[scale=.6]{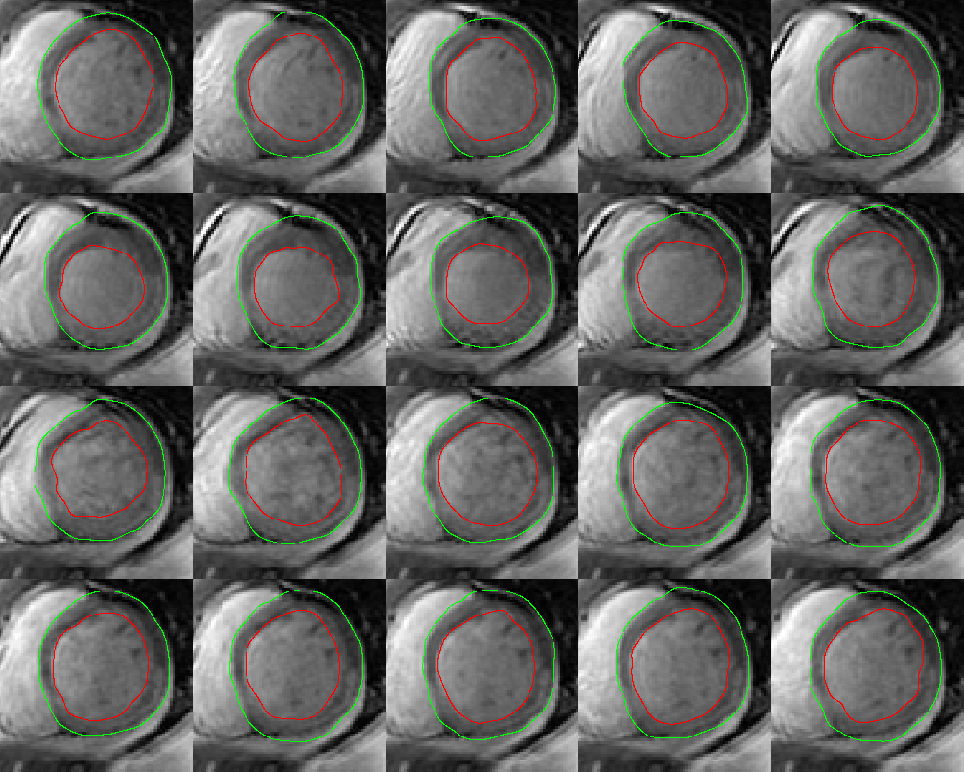}
\caption{Samples of the database with segmented inner (red) and outer (green) boundaries}
\end{figure}

\subsection{Analysis of Volumetric Changes in Left Ventricle}

Analysis of volumetric changes of left ventricle in cardiac cycle can be helpful in
understanding the process of heart muscle's deformation. Also, it helps us study the effects of infarction on the heart's performance.
For this analysis, the data provided for inner contours of left ventricle is used to approximate
the volume. In order to better understand the process and reduce the errors caused
by the differences in the sizes of left ventricle of patients, the computed volumes are normalized with
respect to the initial volume in the beginning of systole. Fig. 7 displays the
normalized graphs of volumetric changes for the three aforementioned patients. 

\begin{figure}
\centering
\includegraphics[scale=.6]{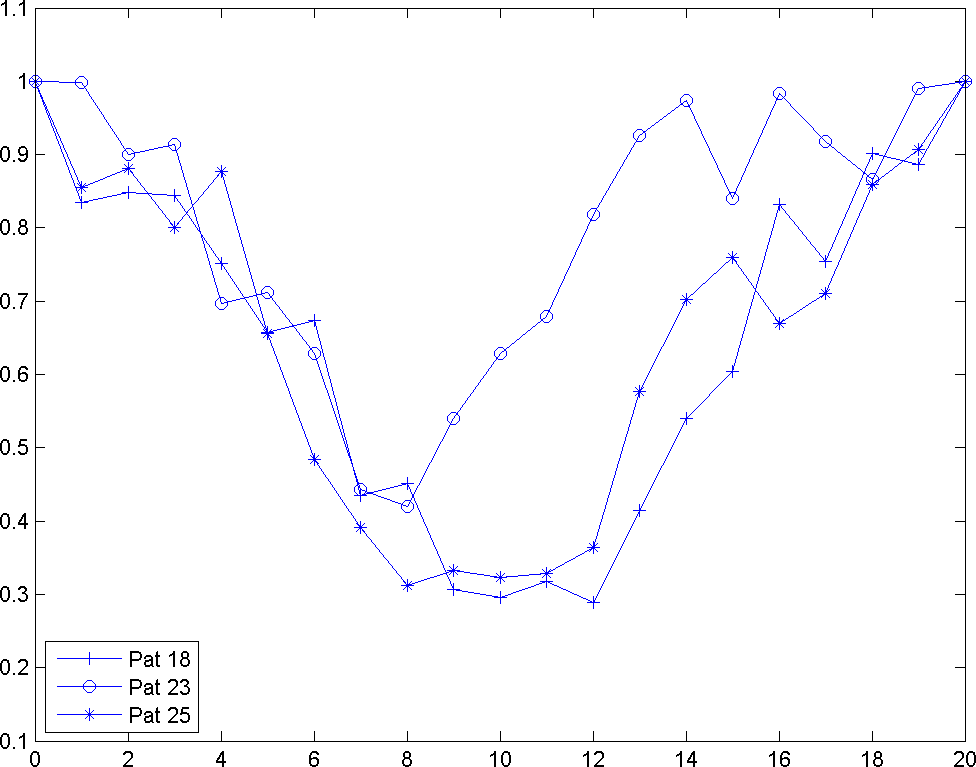}
\caption{Changes in the left ventricle's volume during systole for normal (18 and 25) and MI (23) subjects}
\end{figure}

The first difference that is noticeable comparing these graphs is between frames 0 to 3. In this
region the healthy subjects, 18 and 25, show a small decrease followed by a two frame rest.
There is a delay in patient 23 due to inactive regions in the heart muscle. 
On the other hand in patient 23 the minimum volume is 42\% of the maximum volume which
for healthy subjects this minimum volume is about 30\% (31.66\% for patient 25 and 28.86\%
for patient 18). This is also due to the existence of infraction and the effect of loading
and wasting energy caused by them.

The other difference is that in the unhealthy subject the diastole starts sooner. To be
specific, in 18 and 25 there is a rest region between frames 9-12 and 8-12 respectively. For
23 this region is limited between frames 7-8. It can be inferred that this effect is caused by
the residual stress and inertial effects of inactive regions.
Although the analysis of volumetric changes can help to distinguish between the healthy and
unhealthy heart muscle, it doesn't provide any information about the location of MI.

\subsection{Dynamic Modeling Using Cardiac Images }

In this step MRI images of healthy and unhealthy subjects will be used. From the discussion
in previous section it is obvious that performance and deformation of healthy cardiac
muscles in subjects 18 and 25 are quite identical. Therefore for comparison it is enough to
only consider one of them. It should be mentioned that the more the data for healthy and MI
subjects, the better the results.

Inner and outer contours in MRI images are determined with hand by \cite{23}. Each contour
consists of 32 points. The goal here is to find the one-to-one correspondence between
contour points between frames. The resolution of the data points is increased by interpolation. Another thing to consider here is that the deformation of
heart muscle is not only a concentric radial deformation, but it also contains rotation. On
average, during systole, the cardiac muscle has a 7 degree clockwise rotation which needs to be considered in finding one-to-one correspondence and displacement vectors' computation. The contour points are then down-sampled to reduce the computational complexity. Fig. 8 displays the displacement vectors of one slice of MRI data for subjects 23 (b) and 18 (a) during systole. The existence of MI has a visible effect on the deformation of cardiac muscle. 

\begin{figure}
\centering
\begin{subfigure}[b]{.4\textwidth}
\includegraphics[scale=.4]{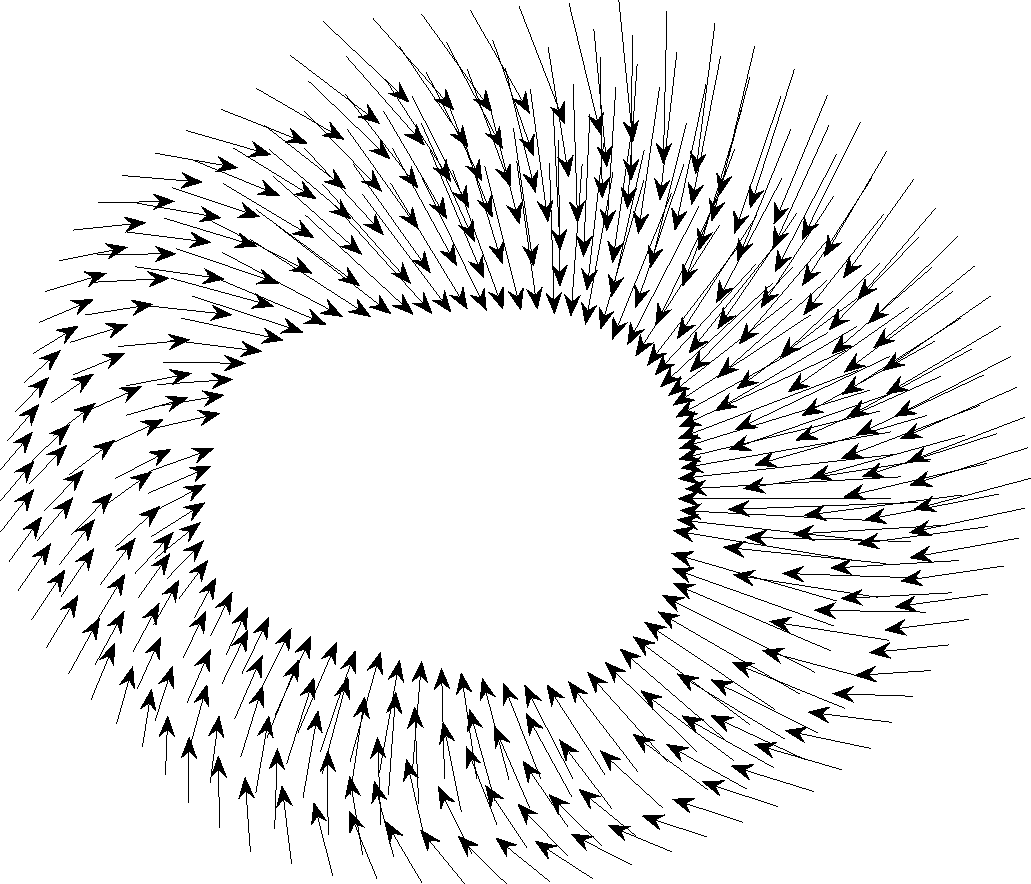}
\caption{}
\end{subfigure}
\begin{subfigure}[b]{0.4\textwidth}
\includegraphics[scale=.4]{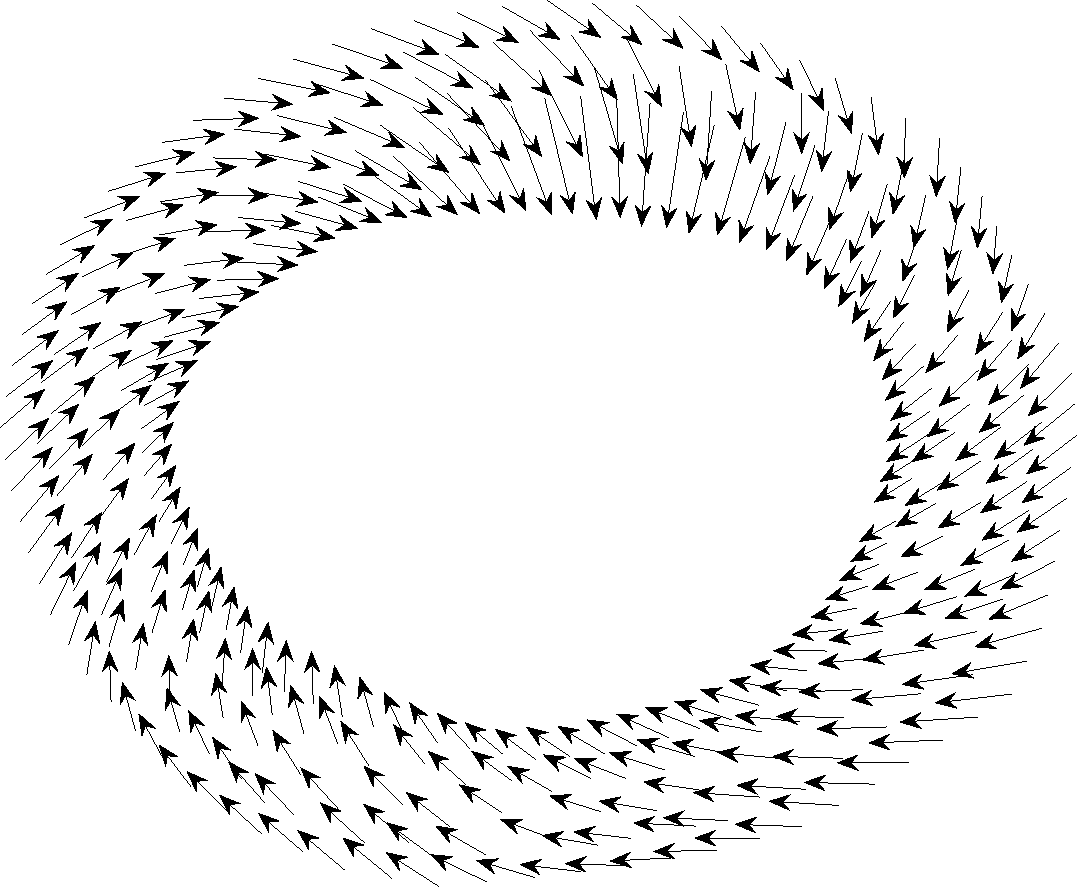}
\caption{•}
\end{subfigure}
\caption{Displacement vectors during systole for one slice of normal (a) and MI (b) subjects}
\end{figure}

Using the displacement vectors, the boundary conditions of the FEM model can be defined.
After applying the boundary conditions and then meshing the domain with a fine triangular mesh, FEM is used to solve the governing equation of deformation. The result of FEM is the displacement values in $x$ and $y$ directions in the deforming body's domain. Fig.9 shows the results of FEM for two slices of subjects 25 and 23 during systole for horizontal and vertical displacements. 

\begin{figure}
\centering
\begin{subfigure}[b]{.45\textwidth}
\includegraphics[scale=.4]{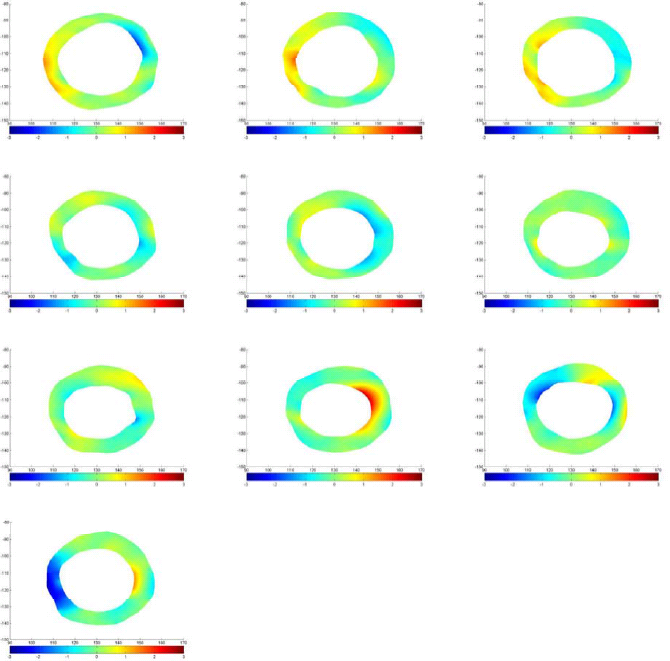}
\caption{range=$[-3, 3]$}
\end{subfigure}
\begin{subfigure}[b]{0.45\textwidth}
\includegraphics[scale=.4]{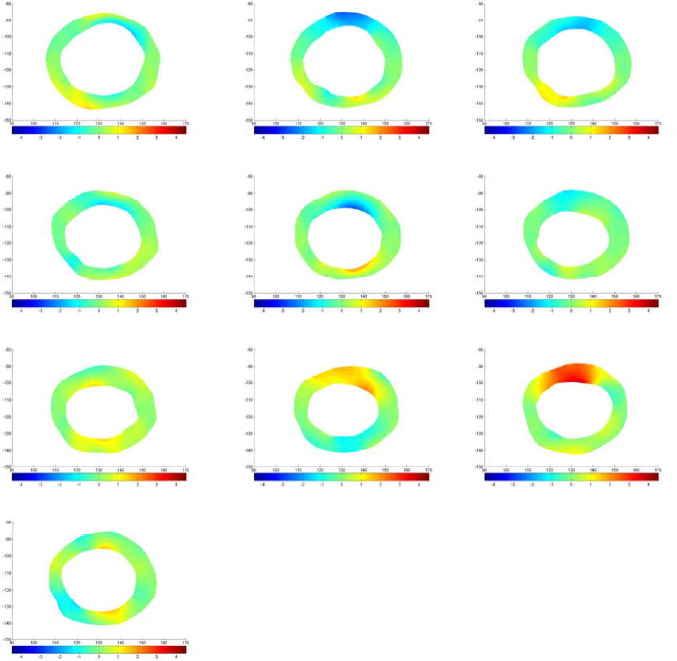}
\caption{range=$[-4, 4]$}
\end{subfigure}

\begin{subfigure}[b]{.45\textwidth}
\includegraphics[scale=.4]{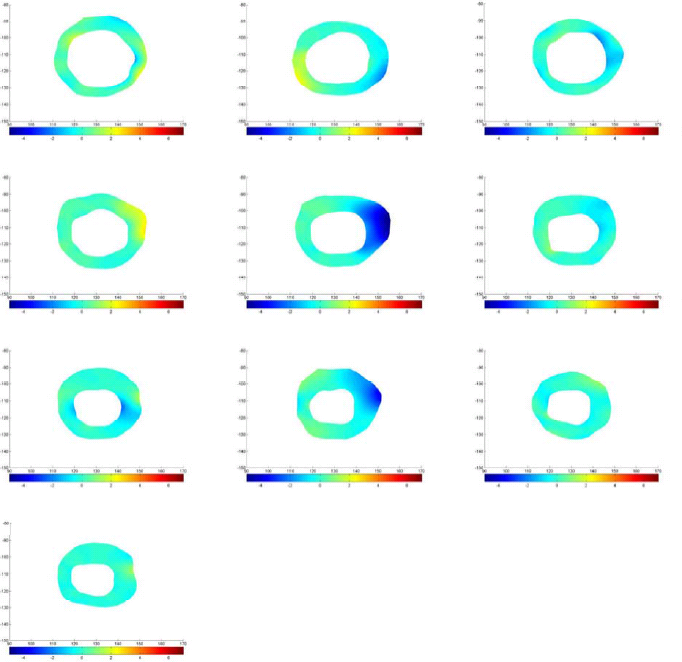}
\caption{range=$[-5, 7]$}
\end{subfigure}
\begin{subfigure}[b]{0.45\textwidth}
\includegraphics[scale=.4]{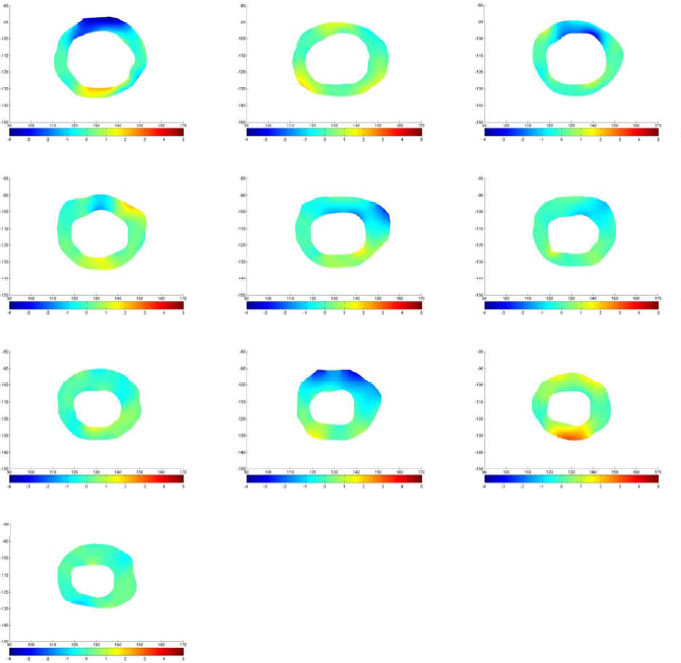}
\caption{range=$[-4, 5]$}
\end{subfigure}
\caption{Horizontal (left) and vertical (right) displacement maps for MI (top) and normal (bottom) subjects during systole}
\end{figure}

Strain elements can be computed using the displacement values . However strain elements only represent the rate of change in deformation with respect to the original configuration in each axis. In case of cardiac muscle analysis it is better if changes in strain is treated in a local basis. On the other hand analyzing 3 strain elements simultaneously can be confusing. Therefore it is better if a unique parameter is defined for analysis. Although there are several different parameters that can be defined, based on the literature in continuum mechanics, here the \textit{effective strain} is used which is defined in general 3D case as \cite{21}: 
\begin{equation}
E_s=\frac{\sqrt{(\varepsilon_x-\varepsilon_y)^2+(\varepsilon_y-\varepsilon_z)^2+(\varepsilon_x-\varepsilon_z)^2+1.5(\gamma_{xy}^2+\gamma_{xz}^2+\gamma_{yz}^2)}}{(1+\nu)\sqrt{2}}
\end{equation}
which in 2D case can be simplified because $\gamma_{xz}=\gamma_{zy}=\varepsilon_z=0$.  Using this parameter we are
able to monitor the changes in strain maps of heart muscle during deformation.
Computing effective strain values for different slices of cardiac muscle, these values are used
for comparison between healthy and MI subjects. Fig. 10 represent the results for
first slice of healthy subjects, 18 and 25, and MI subject 23 during the cardiac cycle.

\begin{figure}
\centering
\begin{subfigure}[b]{.45\textwidth}
\includegraphics[scale=.5]{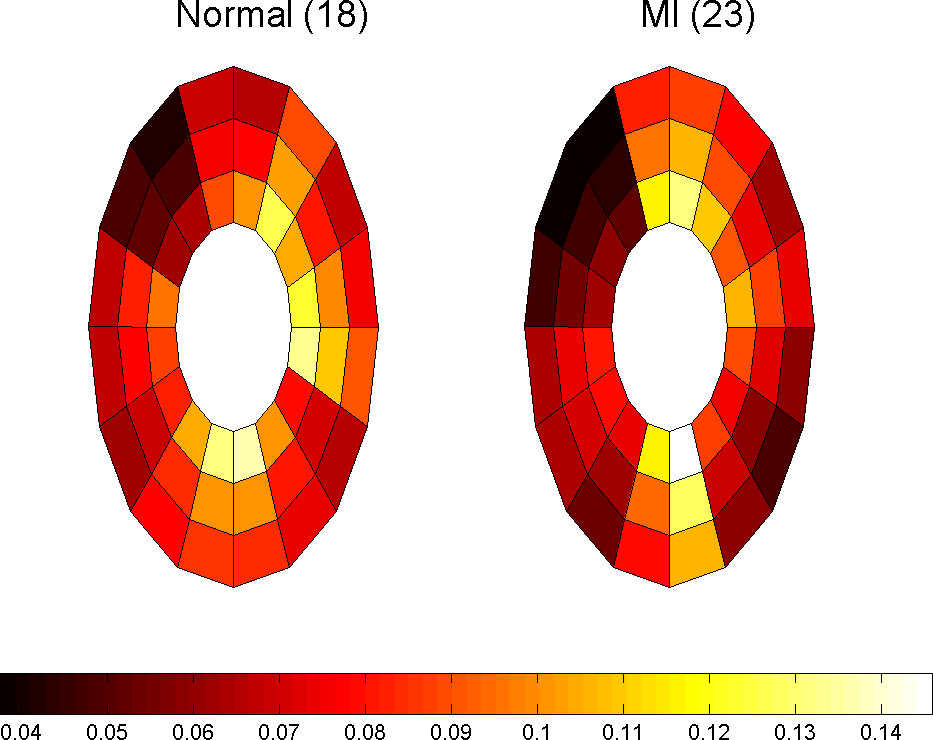}
\caption{}
\end{subfigure}

\begin{subfigure}[b]{0.45\textwidth}
\includegraphics[scale=.5]{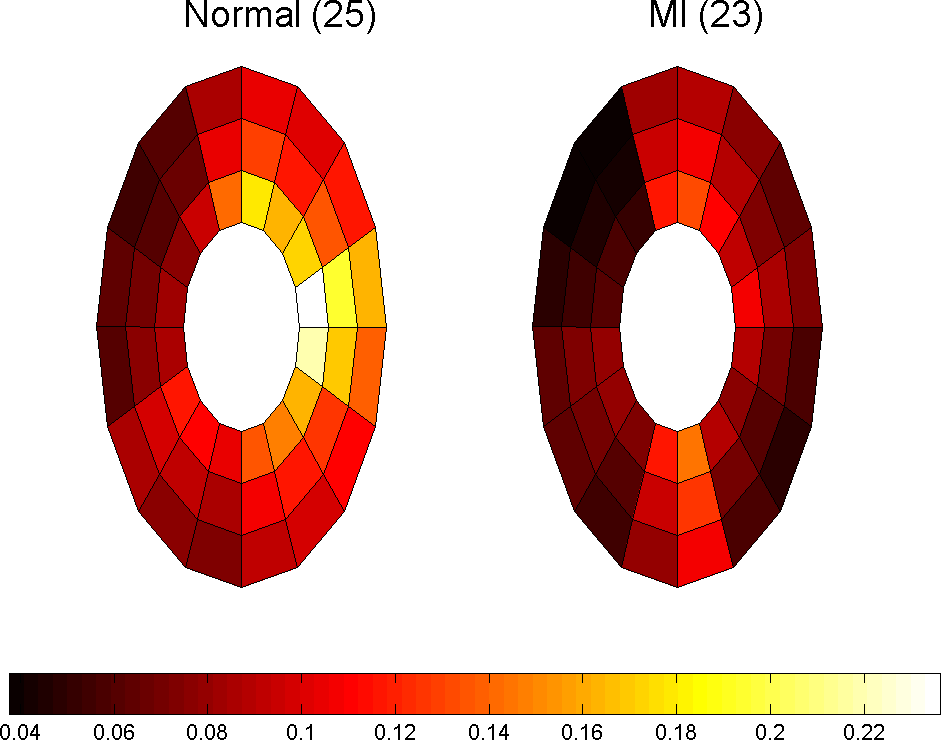}
\caption{}
\end{subfigure}
\caption{Comparison of the effective strain between normal (18 and 25) and MI (23) subject}
\end{figure}

Fig. 11 shows the approximate infarct localization for the MI subject based on the comparison between normal and MI subjects' effective strain during the cardiac cycle. Since the circulatory system of heart is determined very well, it is possible to make reliable conclusions about the problem that caused MI based on expert diagnosis and using these comparisons.  

\begin{figure}
\centering
\includegraphics[scale=.5]{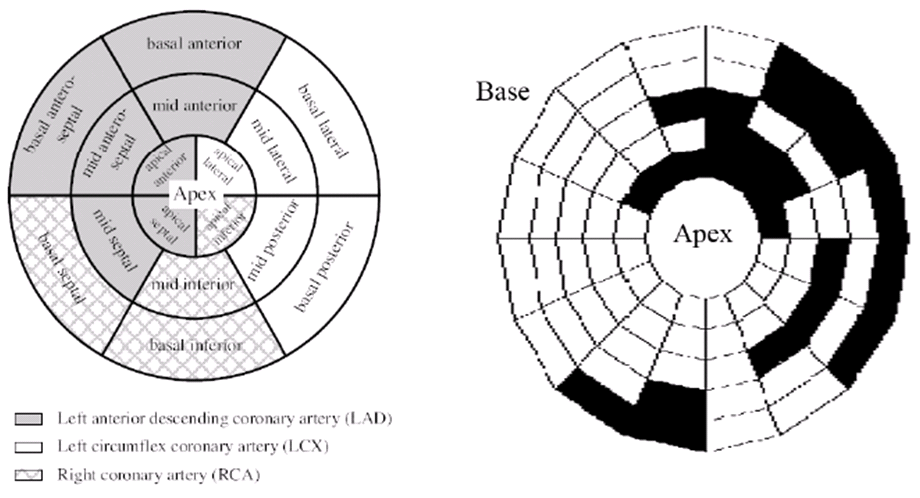}
\caption{Approximate infarct localization for the MI subject (23)}
\end{figure}

\section{Conclusion}
Comparing the results of FEM analysis for both healthy and unhealthy subjects reveals the
effects of infarction in cardiac muscles. Based on the domain and size of infarction, these effects can represent
themselves differently. In this research a consistent model for FEM analysis of cardiac
muscle in case of myocardial infarction is proposed. 
The method is based on defining the displacement vectors for every contour point during
cardiac cycle and then solving the governing equation of deformation with FEM. For a more
complete analysis of the results, a database with more patients is needed. Here only the results for 2 healthy and one MI subjects are compared. But even based on
this simple comparison, the process of MI and its effects on cardiac performance can be
observed very well. As for more sophisticated methods for image processing, use of interpolation or slice interpolation methods \cite{24, 25} can be considered in order to increase the resolution of the MRI images to achieve a better resolution and therefore better 3D model of the heart muscle. Using mesh-based registration techniques like the one proposed in \cite{26} not only reduces the computational complexity of the tracking procedure, but also provides a framework for integrating the process of tracking of the heart muscle with computation of FEM-based solution. Use of noise reduction techniques \cite{27,28} as well as automatic segmentation methods \cite{29} can also be considered for developing more robust and user friendly frameworks with higher reliability.


\begin{thebibliography}{99}\label{ref:ref}

\bibitem{1} Huang, Wen-Chen, and Dmitry B. Goldgof. "Adaptive-size meshes for rigid and nonrigid shape analysis and synthesis." Pattern Analysis and Machine Intelligence, IEEE Transactions on 15.6 (1993): 611-616.

\bibitem {2} McInerney, Tim, and Demetri Terzopoulos. "A dynamic finite element surface model for segmentation and tracking in multidimensional medical images with application to cardiac 4D image analysis." Computerized Medical Imaging and Graphics 19.1 (1995): 69-83.

\bibitem {3} Faber, Tracy L., et al. "Three-dimensional displays of left ventricular epicardial surface from standard cardiac SPECT perfusion quantification techniques." Journal of nuclear medicine: official publication, Society of Nuclear Medicine 36.4 (1995): 697-703.

\bibitem {4} Ranganath, Surendra. "Contour extraction from cardiac MRI studies using snakes." Medical Imaging, IEEE Transactions on 14.2 (1995): 328-338.

\bibitem {5} Nastar, Chahab, and Nicholas Ayache. "Frequency-based nonrigid motion analysis: Application to four dimensional medical images." Pattern Analysis and Machine Intelligence, IEEE Transactions on 18.11 (1996): 1067-1079.

\bibitem {6} Amini, Amir A., Rupert W. Curwen, and John C. Gore. "Snakes and splines for tracking non-rigid heart motion." Computer Vision—ECCV'96. Springer Berlin Heidelberg, 1996. 249-261.

\bibitem {7} Rueckert, Daniel, and Peter Burger. "Geometrically deformable templates for shape-based segmentation and tracking in cardiac MR images." Energy Minimization Methods in Computer Vision and Pattern Recognition. Springer Berlin Heidelberg, 1997.

\bibitem {8} Amini, Amir A., et al. "Coupled B-snake grids and constrained thin-plate splines for analysis of 2-D tissue deformations from tagged MRI." Medical Imaging, IEEE Transactions on 17.3 (1998): 344-356.

\bibitem {9} Abrishami Moghaddam, H., Y. Maingourd, and J. F. Lerallut. "A deformable model based system for 3D analysis and visualization of left ventricle in MRI cardiac images." RBM-News 19.3 (1997): 81-89. 

\bibitem {10} Shi, Pengcheng, et al. "Point-tracked quantitative analysis of left ventricular surface motion from 3-D image sequences." Medical Imaging, IEEE Transactions on 19.1 (2000): 36-50.

\bibitem {11} Konofagou, E. E., T. Harrigan, and S. Solomon. "Assessment of regional myocardial strain using cardiac elastography: Distinguishing infarcted from non-infarcted myocardium." Ultrasonics Symposium, 2001 IEEE. Vol. 2. IEEE, 2001.

\bibitem {12} Pislaru, Cristina, et al. "Higher myocardial strain rates duringisovolumic relaxation phase than duringejection characterize acutely ischemic myocardium." Journal of the American College of Cardiology 40.8 (2002): 1487-1494.

\bibitem {13} Hu, Zhenhua, Dimitris Metaxas, and Leon Axel. "In vivo strain and stress estimation of the heart left and right ventricles from MRI images." Medical Image Analysis 7.4 (2003): 435-444.

\bibitem {14} Hoffmann, Udo, et al. "Acute Myocardial Infarction: Contrast-enhanced Multi–Detector Row CT in a Porcine Model 1." Radiology 231.3 (2004): 697-701.

\bibitem {15} Heijman, Edwin, et al. "Magnetic resonance imaging of regional cardiac function in the mouse." Magnetic Resonance Materials in Physics, Biology and Medicine 17.3-6 (2004): 170-178.

\bibitem {16} Yan, Andrew T., et al. "Characterization of the peri-infarct zone by contrast-enhanced cardiac magnetic resonance imaging is a powerful predictor of post–myocardial infarction mortality." Circulation 114.1 (2006): 32-39.

\bibitem {17} Le Rolle, Virginie, et al. "Model-based analysis of myocardial strain data acquired by tissue Doppler imaging." Artificial intelligence in medicine 44.3 (2008): 201-219.

\bibitem {18} Li, Yinbo, et al. "A four-dimensional model-based method for assessing cardiac contractile dyssynchrony in mice." Ultrasonics Symposium, 2008. IUS 2008. IEEE. IEEE, 2008.

\bibitem {19} Garcia-Barnes, Jaume, et al. "Regional motion patterns for the Left Ventricle function assessment." Pattern Recognition, 2008. ICPR 2008. 19th International Conference on. IEEE, 2008.

\bibitem {20} Baghaie, Ahmadreza, and H. Abrishami Moghaddam. "A consistent model for cardiac deformation estimation under abnormal ventricular muscle conditions." World Congress on Medical Physics and Biomedical Engineering, September 7-12, 2009, Munich, Germany. Springer Berlin Heidelberg, 2010.

\bibitem {21} Popov, Egor Paul, and Toader A. Balan. Engineering mechanics of solids. Vol. 2. Englewood Cliffs, NJ: Prentice Hall, 1990.

\bibitem {22} Cook, Robert D. Concepts and applications of finite element analysis. John Wiley \& Sons, 2007.

\bibitem {23} Andreopoulos, Alexander, and John K. Tsotsos. "Efficient and generalizable statistical models of shape and appearance for analysis of cardiac MRI." Medical Image Analysis 12.3 (2008): 335-357.

\bibitem{24} Baghaie, Ahmadreza, and Zeyun Yu. "Structure tensor based image interpolation method." AEU-International Journal of Electronics and Communications 69.2 (2015): 515-522.

\bibitem{25} Baghaie, Ahmadreza, and Zeyun Yu. "Curvature-Based Registration for Slice Interpolation of Medical Images." Computational Modeling of Objects Presented in Images. Fundamentals, Methods, and Applications. Springer International Publishing, 2014. 69-80.

\bibitem{26} Baghaie, Ahmadreza, Zeyun Yu, and Roshan M. D’souza. "Fast Mesh-Based Medical Image Registration." Advances in Visual Computing. Springer International Publishing, 2014. 1-10.

\bibitem{27} Baghaie, Ahmadreza, Roshan M. D'souza, and Zeyun Yu. "Sparse And Low Rank Decomposition Based Batch Image Alignment for Speckle Reduction of retinal OCT Images." arXiv preprint arXiv:1411.4033 (2014).

\bibitem{28} Baghaie, Ahmadreza, Roshan M. D'souza, and Zeyun Yu. "Application of Independent Component Analysis Techniques in Speckle Noise Reduction of Single-Shot Retinal OCT Images." arXiv preprint arXiv:1502.05742 (2015).

\bibitem{29} Baghaie, Ahmadreza, Roshan M. D'souza, and Zeyun Yu. "State-of-the-Art in Retinal Optical Coherence Tomography Image Analysis." arXiv preprint arXiv:1411.0740 (2014).


\end{thebibliography}
\end{document}